\documentstyle[12pt]{article}
\newcommand{\Tr}[1]{\,{\rm Tr}\,#1\,}

\begin{document}
\title{
\begin{flushright}
{\small UTCCP-P-43
}
\end{flushright}
\vspace{0.5cm}
Lattice QCD with Exponentially Small Chirality Breaking.}
\author{A.A.Slavnov \thanks{on leave from Steklov Mathematical
Institute, Gubkina st. 8, GSP-1, 117966 Moscow, Russia} \\Center for
Computational Physics, University of Tsukuba, \\ Tsukuba, Ibaraki 305,
Japan} \maketitle

\begin{abstract}

A new multifermion formulation of lattice QCD is proposed. The model
is free of spectrum doubling and preserves all nonanomalous chiral
symmetries up to exponentially small corrections. It is argued that a
small number of fermion fields may provide a good approximation making
computer simulations feasible. 

\end{abstract}

\section{Introduction}

It was believed for a long time that any local regularization of a
theory which posesses a chiral symmetry breaks a manifest chiral
invariance. In the context of lattice models this statement is the
essence of the Nielsen-Ninomiya ``no-go'' theorem \cite{NiN}. However in
our paper \cite{FS1} it was shown that anomaly free models, such as
the Standard Model, do allow a local chiral invariant regularization
if one introduces the regularized action containing infinite number of
fermion fields. Application of this idea to lattice theories lead to
the models which have no spectrum doubling and posess exact chiral
invariance in the continuum limit \cite{FS2},
\cite{S1}. The method was also checked by nonperturbative
simulations of some $2d$ models \cite{SZ}. R.Narayanan and H.Neuberger
showed \cite{NN} that this approach is closely related to the five
dimensional domain wall model proposed by D.Kaplan
\cite{K}. Introduction of an infinite series of fermion fields may be
considered as a discretization of the fifth dimension.

Although originally it might seem that considering an infinite series
of fermions is too high price, by now it is widely accepted that it is
the only way to preserve the exact chiral symmetry without tuning
additional parameters. Many present approaches like overlapp
formalism \cite{NN2} or different versions of domain wall fermions
\cite{Sh}, \cite{ShF} use this idea.

The main question is whether these approaches allow efficient
nonperturbative calculations. Considering realistic chiral gauge
models on the lattice seems at present to be too hard for computer
simulations. On the top of the usual problem of simulating fermion
determinants there is an additional difficulty, as a chiral
determinant being complex does not allow straightforward Monte-Carlo
simulations. So it seems reasonable to start with the study of
vectorial models like QCD and to try to deal with the breaking of the
global chiral symmetry.

A naive introduction of the Wilson term in QCD to remove the
degeneracy of fermion spectrum breaks global chiral invariance of the
action leading to appearance of quark mass counterterms as well as
order $a$ chirality breaking corrections. Apart from esthetical
objections it raises  practical questions for numerical simulations.
Whereas the Abelian chiral symmetry is indeed broken by the triangle
anomaly, the $SU(2)$ chiral invariance is not affected by anomalies
and is believed to be broken spontaneously. Explicit breaking of the
$SU(2)$ symmetry by the Wilson term leads to a nonzero mass of $
\pi$-meson which is supposed to be a pceudogoldstone boson.

To supress these effects one may try to modify the Wilson action by
adding the terms killing the order $a$ corrections \cite{ShW}. It
partially solves the problem, but the modified action makes computer
simulations more difficult. Extensive study of this approach was
carried out in the papers (see \cite{L1} and references within).

It seems preferrable to use some approach which preserves, at least
approximately, the global chiral symmetry of the theory. One
possibility, which is discussed now, is to use the actions, satisfying
the Ginzparg-Wilson relation \cite{GW}. It was shown, that if the
model satisfies this relation, a number of results which one expects
in a chiral invariant theory follows \cite{Has}, \cite{Nid}. Moreover
 in this case some modified chiral symmetry is
present \cite{L2}. Examples of such actions were given in the papers
\cite{N1}, \cite{L2}. They are typically nonlinear and
formally nonlocal functionals, but nonlocal effects are damped
exponentially. Some qualitative confirmation of the relevance of this
approach for low energy QCD was presented in the paper \cite{Ch}.

The multifermion models like truncated overlapp \cite{N2} or domain
wall fermions \cite{ShF} were also applied to the study of QCD and
first numerical simulations \cite{BS}, \cite{B} gave promising results. The
chirality breaking effects, in particular fermion mass
renormalization, were also discussed \cite{Ao}, \cite{Kik}.

In the present paper I propose a new version of a multifermion
formulation of QCD. The model has no species doubling, does not
require any fine tuning to calculate arbitrary gauge invariant
amplitudes and does not contain any dimensional parameter except for a
lattice spacing. For a finite lattice spacing all chirality breaking
effects are supressed exponentially. Although the model includes an
infinite number of fermion species, the convergence of the
corresponding series is very fast, and one can hope that it may be
truncated by a small finite number of terms. 

The paper is organized as follows. In the Section 1 we formulate a
general idea of the method and prove that our effective action
produces the fermion determinant differing from the gauge invariant QCD
determinant by exponentially small terms. In the second section we
study the chiral property of the model and demonstrate the absence of
a perturbative quark mass renormalization as well as exponential
supression of all chirality breaking terms. In the Section 3 it is
shown that all nonanomalous chiral currents are conserved whereas the
usual Abelian anomaly is present. In Section 4 we discuss
possible applications of the model to numerical simulations.

\section{Representation of the QCD quark determinant in the
multifermion model.}

To give an idea of the method we start with the continuum model
describing a gauge invariant interaction of two generations of spinor
fields $ \chi_n, \quad n=1 \ldots \infty$ with the masses proportional
to the number $n$, $m_n=mn$. The following equality holds
\begin{equation}
 \int \exp \{- \sum_{k=1,2} \sum_{n=1}^{ \infty} \int
\bar{\chi}^k_n(x)( \hat{D}+mn) \chi^k_n(x) dx \} d \bar{\chi}^k_n d \chi^k_n=
\label{1}
\end{equation}
$$
 \int \exp \{- \sum_{n=1}^{ \infty} \int \bar{\chi}_n(x)(-D^2+m^2n^2)
\chi_n(x) dx \} d \bar{\chi}_n d \chi_n=
$$
$$
C \prod_i \coth( \frac{ \pi D_i}{2m})D_i
$$
Here $ \chi_n$ are spinor fields with the Grassmanian parity $(-1)^n$,
$D$ is a QCD Dirac operator and $D_i$ are it's eigenvalues. $C$ is a
field independent constant.

To derive this equation we note that the integral over $ \chi$ is
equal to

\begin{equation}
I= \prod_{n=1}^{\infty} \det(-D^2+m^2n^2)^{(-1)^{n+1}}
\label{2}
\end{equation}
This equation may be rewritten as follows
\begin{equation}
I= \prod_i \exp \{ \sum_{n=1}^{\infty} \ln(D_i^2+m^2n^2)(-1)^{n+1} \}
\label{3}
\end{equation}
Summation in the exponent (\ref{3}) may be done explicitely, giving
the result
\begin{equation}
\sum_{n=1}^{\infty} \ln(D_i^2+m^2n^2)(-1)^{n+1}= \ln( \coth
\frac{\pi|D_i|}{2m})+ \ln|D_i|
\label{4}
\end{equation}
This proves the eq.( \ref{1}).

It follows from the eq.(\ref{1}) that up to a nonessential constant factor
\begin{equation}
\lim_{m \rightarrow 0} I= \det(D)
\label{5}
\end{equation} 
and
\begin{equation}
\lim_{m \rightarrow \infty}I=1
\label{6}
\end{equation}
These equations strongly suggest that if we apply analogous procedure to
lattice QCD, and instead of the mass term introduce the Wilson term
$W$, we get a theory which for small values of $W$ produces the QCD
massless quark determinant and for large values of $W$ (doublers
region) is trivial. As follows from eq.(\ref{4}) the doublers
contribution is supressed by the factor $ \exp \{- \frac{ \pi D}{2m}
\}$, so we expect that all chirality breaking effects are also supressed
exponentially. Below we justify this conjecture.

Note that in distinction of our previous approach \cite{FS2},
\cite{S1} as well as different versions of domain wall fermions, which
include heavy states in the domain $p \sim 0$, in the present version
only massless states are present in this region, and heavy states are
confined to the doublers region $p \sim \pi a^{-1}$. No additional
mass parameters appear in the theory.

Our claim is that the QCD determinant may be presented as the path
integral of the exponent of the following action:
\begin{equation}
I= \sum_{x, \mu} \sum_{n=- \infty, n\neq 0}^{\infty}- 
\bar{ \psi}^n(x) \frac{1}{2}[ \gamma_{\mu}(D_{\mu}+D_{\mu}^*)- n \kappa
aD_{\mu}^*D_{\mu}] \psi_n(x) 
\label{7}
\end{equation}
\begin{equation}
D_{\mu}= \frac{1}{a}[U_{\mu}(x) \psi(x+a_{\mu})- \psi(x)]
\label{7a}
\end{equation}
Here the first term represents the chiral invariant part of the action
and the second term is the Wilson term multiplied by $n$. In this equation $\kappa$ is a dimensionless
parameter, which we choose in the interval $0< \kappa <1$. The fields
$ \psi^n$ have a Grassmanian parity $(-1)^n$.

In distinction of the eq.(\ref{1}) where we considered two
generations of $\chi^k_n$ fields $(k=1,2)$, and $1 \leq n \leq \infty$, in
eq.( \ref{7}) we introduced one generation, but allow $n$ to change in
the interval $- \infty \leq n \leq \infty, \quad n \neq 0$. For a zero
bare quark mass it makes no difference, but , as it was pointed out to
me by A.Ukawa, for nonzero quark mass the choice $- \infty \leq n \leq
\infty, \quad n \neq 0$ is more convinient. 

Let us consider a fermion loop diagram with $L$ vertices, which may be
presented by the following integral
 \begin{equation}
\Pi_L(q_1 \ldots q_{L-1})= \int_{- \pi a^{-1}}^{ \pi a^{-1}}d^4p
\sum_{n=- \infty, n \neq 0}^{\infty}(-1)^n \Tr[G(p) \Gamma_i(p,q_1)G(p+q_1)
\Gamma_j(p,q_2) \ldots ]=
 \label{8}
 \end{equation}
 $$
=2 \int_{- \pi a^{-1}}^{ \pi a^{-1}} \sum_{n=1}^{\infty}(-1)^n
\frac{F_0(p,q)+n^2F_1(p,q)+ \ldots}{(s^2+m^2n^2)(s_1^2+m_1^2n^2) \ldots}
 $$
 Here $ \Gamma_i$ denotes the interaction vertices and $q_l$ is a
total momentum entering the corresponding vertex.
 The fermion propagators $G(p)$ look as follows
 \begin{equation}
G_n(p)= \frac{ i \hat{s}+mn}{s^2+m^2n^2}
 \label{9}
 \end{equation}
 \begin{equation}
s_{\mu}^l=a^{-1} \sin[(p+q^1+ \ldots +q^l)_{\mu}a]
 \label{10}
 \end{equation}
 \begin{equation}
m^l= \kappa a^{-1} \sum_{\mu}(1- \cos(p+q^1+ \ldots +q^l)_{\mu})
 \label{11}
 \end{equation}
The eq.(\ref{8}) may be rewritten in the following form
 \begin{equation}
 \Pi_L=2 \int_{- \pi a^{-1}}^{\pi a^{-1}}d^4p \sum_{l=0}^{L-1}
\sum_{n=1}^{\infty}(-1)^n \frac{A_l(p,q)}{s_l^2+m_l^2n^2}
 \label{12}
 \end{equation}
 Here $A_l$ are some functions of $p,q$, which do not depend on $n$.
They satisfy the equations
\begin{equation}
\sum_lA_l \prod_{i=0, i\neq l}^{L-1}s_i^2=F_0(p,q)
\label{13}
\end{equation}
$$
 \ldots
$$
$$
 \sum_l A_l\prod_{i \neq l}m_i^2=0
$$
The summation over $n$ can be done as before, giving the result  
 \begin{equation}
 \Pi_L= \int_{- \pi a^{-1}}^{\pi a^{-1}}d^4p[- \sum_{l=0}^{L-1} \frac
{ \pi A_l(p,q)}{m_l \sqrt{s_l^2} \sinh( \pi \sqrt{s_l^2}m_l^{-1})}+ \frac{A_l(p,q)}{s_l^2}]
 \label{14}
 \end{equation}
One sees that for small $p+q_1+ \ldots +q_l$ the first term is damped
exponentially and the second one describes the amplitude for massless
quarks. For $p+q_1+ \ldots +q_l \sim \pi a^{-1}$ the first term is 
 \begin{equation}
 \sim - \frac{A_l(p,q)}{s_l^2}
  \label{15}
 \end{equation}
and exactly compensates the second one providing the doublers
supression.

Now we shall study eqs. (\ref{8}- \ref{14}) in more details and prove
that the amplitudes (\ref{8}) coincide with the gauge invariant QCD
amplitudes up to the terms which decrease exponentially when $a
\rightarrow 0$.

Firstly we consider the case when all momenta $q_l$ are external. Then
we can assume that
 \begin{equation}
|q_l|< \epsilon a^{-1}, \quad  \epsilon<1
 \label{16}
 \end{equation}
The integration domain in the eq.(\ref{8}) can be separated into three
parts: 
 \begin{equation}
V_1: |p|< \epsilon a^{-1}; \quad V_2: L \epsilon a^{-1} \leq |p| \leq
( \pi-L \epsilon)a^{-1};
 \label{17}
 \end{equation}
$$
 V_3: ( \pi-L \epsilon)a^{-1} \leq |p| \leq \pi a^{-1}
$$
In the region $V_2$ the function 
\begin{equation}
s_l^{-2} \simeq (\sum_{\mu} \sin(p_{\mu}a)(1+ \cot(p_{\mu}a)(q_1+ \ldots
+q_l)a)))^{-2}
\label{18}
\end{equation}
may be expanded in terms of $q_la$. It is easy to see that the the
functions $A_l$ also may be expanded in terms of $q_la$. Therefore
integrating the second term in the eq.(\ref{14}) over $p$ one gets a
local polynomial in $q_l$. Analogous reasonings show that the first
term in the eq.(\ref{14}) also contributes a local polynomial in
$q_l$.

In the domain $V_3$ we cannot use the expansion (\ref{18}) as in this
region $ \sin(p_{\mu}a) \sim 0$. However the complete r.h.s. of the
eq.(\ref{14}) still allows the expansion in terms of $q_la$. Indeed,
in this region $s_l^2<m_l^2$, and the r.h.s. of the eq(\ref{14}) may
be presented as follows
\begin{equation}
\Pi_L= \int_{- \pi a^{-1}}^{ \pi a^{-1}}d^4p \sum_{l=0}^{L-1}A_l 
\{ \frac{1}{s_l^2}- \frac{1}{s_l^2}(1- \frac{ \pi^2 s_l^2}{6
\pi^2m_l^2}+O( \frac{s^4}{m^4})) \}
\label{19}
\end{equation}
The terms $ \sim s_l^{-2}$ cancel and the remaining terms may be
expanded in terms of $q_la$.

In the region $V_1$ we use the representation (\ref{14}) for the
amplitude (\ref{8}). In this region $ |a(p-q_l)|<2 \epsilon$ and therefore
 \begin{equation}
\sum_{\mu} \sin^2[a(p-q_l)_{\mu}]>[ \sum_{\mu}1- \cos(p-q_l)_{\mu}a]^2
 \label{20}
 \end{equation}
Hence the first term in the eq.(\ref{14}) is supressed exponentially
by the factor 
 \begin{equation}
\frac{1}{\sinh(\pi \sqrt{s_l^2}m_l^{-1})} \sim \exp \{- \frac{\pi}{\epsilon \kappa}\}
\label{21}
\end{equation}
The second term gives
 \begin{equation}
\int_{- \epsilon a^{-1}}^{\epsilon a^{-1}}d^4p \sum_{l=0}^{L-1}
\frac{A_l(p,q)}{s_l^2}= \int_{\epsilon a^{-1}}^{\epsilon a^{-1}}d^4p
\frac{F_0(p,q)}{s_0^2 \ldots s_{L-1}^2}
 \label{22}
 \end{equation}
where the eq.(\ref{13}) was used. This is exactly the expression which
one would get in the chiral invariant theory with no Wilson term. In
the limit $a \rightarrow 0$ this term reproduces the diagrams of the
continuum QCD. It is worthwhile to notice that
if only this term were present, the gauge invariance could be broken,
as cutting the Brillouin zone is not a gauge invariant procedure. In
our construction the gauge invariance is preserved at any stage of
calculations. Possible noninvariant terms arising from cutting the
Brillouin zone in the eq.(\ref{22}) are compensated by local
counterterms generated by the integral over $V_2, V_3$. It would be
wrong just to drop the integration over $V_2, V_3$. The local
counterterms coming from these integrals are crucial for maintaining
the gauge invariance of the amplitudes. So we proved
that for one loop diagrams, when all $q_l$ are external momenta, $q_l<
\epsilon a^{-1}$, the amplitude (\ref{8}) differs from the gauge
invariant QCD amplitude by exponentially small terms. The value of the
correction term for the fixed $a$ depends on external momenta $q$. In
particular for $q=0$ it is zero for any $a$. The rate of the exponential
damping depends also on the parameter $\kappa$, and one may have a
temptation to take $\kappa$ very close to zero. It is clear however
that if one puts $ \kappa=0$, the correction term indeed disappear, but
the doubler states reappear. To determine possible values of $\kappa$
we note that considering the contribution of the region $ V_3$ we
used the expansion in terms of $qa$. It is easy to see that for $p
\sim \pi a^{-1}$ the effective expansion parameter is
$\frac{qa}{\kappa}$. So the expansion to make sense we need
$\kappa>qa$. In particular we may take $\kappa= \epsilon$.

In the next section we consider the multiloop diagrams and prove that
they also reproduce chiral invariant QCD amplitudes up to
exponentially small corrections. To do that we have to study more
closely the global chiral invariance of our model.

 \section{Global chiral invariance of the model.}
 
In this section we study the invariance of our model under global
chiral transformations. We show, that as expected in a chiral
invariant theory, a perturbative quark mass renormalization is absent.
All chirality breaking effects are supressed exponentially.
The divergency of any amplitude including an $SU(2)$ chiral current
vertex, or any other chiral current, corresponding to an anomaly free
group is zero up to exponentially small corrections. At the same time
the divergency of the Abelian axial current is not zero, producing the
usual chiral anomaly.

To give an idea of the mechanism which provides the supression of
chirality breaking effects we firstly consider a simplified case. All
chirality breaking effects arise due to the presence of the Wilson
term and therefore vanish when $n=0$. Let us suppose that some
chirality breaking term, for example the fermion mass renormalization,
is simply proportional to $n$: $ \delta m_n=g^2M(g)n$.

Introducing such counterterm  is equvalent to shifting the Wilson term by
$g^2Mn$. To get the expression for the QCD determinant we still may
use the eq.s( \ref{8}- \ref{14}), changing everywhere $m_l$ by
$m_l+g^2M$. It is sufficient to consider the case when $|q_l|< \epsilon
a^{-1}$ and the integration domain in the eq.( \ref{14}) is restricted
to $( - \epsilon a^{-1}, \epsilon a^{-1})$, as it was shown above
that larger values of $p$ produce only local counterterms needed to
maintain the gauge invariance.
The expression for the QCD amplitude acquires the form:
\begin{equation}
\Pi_L= \int_{- \epsilon a^{-1}}^{ \epsilon a^{-1}}d^4p
[ \frac{A_l(p,q)}{s_l^2}-
\label{24}
\end{equation}
$$
- \sum_{l=0}^{L-1} \frac{ \pi A_l(p,q)}{(m_l+g^2M) \sqrt{s_l^2} \sinh
( \pi \sqrt{s_l^2}(m_l+g^2M)^{-1})}]
$$
The sum of the first terms in the eq.(\ref{24}) due to eq.(\ref{13})
is equal to 
\begin{equation}
\int_{- \epsilon a^{-1}}^{ \epsilon a^{-1}}d^4p \frac{F_0(p,q)}{s_0^2
\ldots s_{L-1}^2}
\label{25}
\end{equation}
which coincides with the result obtained before for the model without
mass counterterm insertions. Expanding the second term into Taylor
series over $g$ one sees that all the coefficients are exponentially
damped, being proportional to $ \exp\{- \pi \sqrt{s_l^2}m_l^{-1}\} \leq
\exp \{- \pi( \kappa \epsilon)^{-1}\}$. So for the simplified case we
proved our statement.

However in general chirality breaking terms may have more complicated
dependence on $n$. Below we shall show that our statement remains true
in a general case as well, although the proof is more involved.

To prove the chiral invariance of the determinant it is sufficient to
show that it's derivative with respect to $ \kappa$ is proportional to
some chiral invariant amplitude. Note that this derivative is not
necessarily zero as chiral invariant counterterms may also depend on $
\kappa$. We demonstrate that it is indeed true up to exponentially
small terms.

We study the following object
\begin{equation}
I= \frac{ \partial}{ \partial \kappa}(\sum_{n=- \infty, n \neq
0}^{\infty}(-1)^n \Tr \ln D_n)=
\label{26}
\end{equation}
$$
= \sum_n(-1)^n \Tr(nW(D+n \kappa W)^{-1})
$$
Here $W$ is the usual Wilson term. This equation may be rewritten as
follows
\begin{equation}
I= \sum_n (-1)^n \Tr(nWG_n)
\label{27}
\end{equation}
wher $G_n$ is the Green function of the Wilson-Dirac operator in the
external gauge field. This function satisfies the chiral Ward
identities which in the continuum notations may be written as follows
\begin{equation}
G_n(x,y) \gamma_5+ \gamma_5 G_n(x,y)=2in \kappa \int G_n(x,z)
\gamma_5 W(z)G_n(z,y)dz
\label{28}
\end{equation}
\begin{equation}
 \gamma_5G_n \gamma_5=-G_{-n}
\label{29}
\end{equation}

The Green function $G_n$ may be presented as a sum $G_n=G_n^s+G_n^b$
where $G_n^s$ is a chiral symmetric part which anticommutes with
$\gamma_5$ and is even in $n$. $G_n^b$ is a chirality breaking part
which commutes with $\gamma_5$ and is odd in $n$. It follows from
eq(\ref{28}) that
\begin{equation}
G_n^b(x,y)=in \kappa \gamma_5 \int G_n(x,z) \gamma_5 W(z)G_n(z,y)dz
\label{30}
\end{equation}
Using these properties of $G_n$ one can rewrite the eq(\ref{27}) in
the form
\begin{equation}
I= \sum_{n=- \infty, n \neq 0}^{ \infty}(-1)^nn^2i \kappa 
\int \Tr(W(x) \gamma_5 G_n(x,z) \gamma_5 W(z)G_n(z,x))dxdz
\label{31}
\end{equation}
The eq(\ref{31}) is written for one loop diagrams. To get the multiloop
diagrams one has to integrate it with the exponent of the gluon
action. In terms of Feynman diagrams it is represented by the fermion
loops with two insertions of the Wilson term $W$.

We shall start with the diagrams including one fermion loop and
arbitrary number of gluon vertices and lines. 

The analytic expression corresponding to the eq(\ref{31}) for such
diagrams may be written as follows
\begin{equation}
I_L=\int \sum_{n=- \infty, n \neq 0}^{\infty}(-1)^nn^2
\frac{(K_0+n^2K_1+ \ldots)F(q_r \ldots q_{L-1})dpdq_r
dq_{L-1}}{(s_0^2+m_0^2n^2) \ldots
(s_{L-1}^2+m_{L-1}^2n^2)}
\label{32}
\end{equation}
Here we wrote the expression for the diagram with $L$ quark- gluon
vertices. The notations are as follows:
\begin{equation}
s_0= \sqrt( \sum_{\mu}s^2_{\mu}(p)), \quad m_0=m(p), 
\label{33}
\end{equation}
$$
 s_l= \sqrt( \sum_{\mu}s^2_{\mu}(p+q_1+ \ldots +q_l)), \quad m_l=m(p+
+q_1+\ldots +q_l)
$$
The momenta $q_1, \ldots q_r$ are external and satisfy the condition
$|q_la|< \epsilon$. The momenta $q_{r+1}, \ldots q_{L-1}$ correspond
to virtual gluons and may acquire arbitrary values in the interval $(-
\frac{\pi}{a}, \frac{\pi}{a})$. The function $F(q_r, \ldots q_{L-1})$
depends only on virtual gluon momenta corresponding to gluon
propagators and vertices.

We use the following decomposition of the integrand
\begin{equation}
\frac{K_0+n^2K_1+ \ldots}{(s_0^2+m_0^2n^2) \ldots (s_r^2+m_r^2n^2)}=
\sum_{l=0}^r \frac{B_l}{s_l^2+m_l^2n^2}
\label{34}
\end{equation}
\begin{equation}
\frac{1}{(s_{r+1}^2+m_{r+1}^2n^2) \ldots (s_{L-1}^2+m_{L-1}^2n^2)}=
\sum_{k=r+1}^{L-1} \frac{A_k}{s_k^2+m_k^2n^2}
\label{35}
\end{equation}
Here $B_l$ and $A_k$ are $n$-independent functions satisfying the
relations
\begin{equation}
 \sum_l B_l \prod_{i \neq l}s_i^2=K_0, \quad \ldots \quad \sum_l B_l
\prod_{i \neq l}m_i^2=0
\label{36}
\end{equation}
\begin{equation}
\sum_kA_k \prod_{i \neq k}s_i^2=1, \quad \ldots \quad \sum_kA_k
\prod_{i \neq k}m_k^2=0
\label{37}
\end{equation}
We assume that all momenta $s_l$ are different. It is not true
for the diagrams with self energy insertions, but it will be clear
from the final result that our conclusions are valid for this case as
well.

Using these decompositions we may rewrite the integrand in
eq(\ref{32}) in the form
\begin{equation}
 \tilde{I}_L= \sum_{n=- \infty, n \neq 1}^{\infty}(-1)^nn^2
( \sum_{l=0}^r \frac{B_l}{s_l^2+m_l^2n^2})( \sum_{k=r+1}^{L-1}
\frac{A_k}{s_k^2+m_k^2n^2})=
\label{38}
\end{equation}
$$
= \sum_n(-1)^n \sum_{l,k} \frac{B_lA_k}{s_k^2m_l^2-s_l^2m_k^2}
( \frac{m_l^2n^2}{s_l^2+m_l^2n^2}- \frac{m_k^2n^2}{s_k^2+m_k^2n^2})
$$
Performing the summation over $n$ one gets
\begin{equation}
 \tilde{I}_L= \sum_{l,k} \frac{B_lA_k}{s_k^2m_l^2-s_l^2m_k^2}[ \frac{s_l^2
\pi}{s_lm_l \sinh( \pi s_lm_l^{-1})}- \frac{s_k^2 \pi}{s_km_k \sinh
( \pi s_km_k^{-1})}]
\label{39}
\end{equation}
All the momenta $s_l, \quad 0 \geq l \geq r$ are small and therefore
the first term gives exponentially small corrections. If some of the
momenta $s_k$ are also small, the corresponding term is also supressed
exponentially. So it is sufficient to consider the eq(\ref{39}) when
all $s_k$ are big: $ \epsilon \leq |a(p+ \ldots q_k)| \leq \pi$.

The remaining terms may be expanded in terms of $m_l^2s_l^{-2}$. (We
recall that $q_la< \epsilon$ and therefore $m_l \ll s_l$.
\begin{equation}
\tilde{I}_L= \sum_{k=r+1}^{L-1} \frac{A_ks_k^2 \pi}{s_km_k^3 \sinh
( \pi s_km_k^{-1})} \sum_{l=0}^r \frac{B_l}{s_l^2}(1+
\frac{m_l^2s_k^2}{s_l^2m_k^2} + \ldots)
\label{40}
\end{equation}
Let us consider the sum over $l$. Due to eq(\ref{35}) the first term
is equal to
\begin{equation}
\sum_{l=0}^r \frac{B_l}{s_l^2}= \frac{K_0}{s_0^2 \ldots
s_r^2}=s_{r+1}^2 \ldots s_{L-1}^2 I_L^0
\label{41}
\end{equation}
where 
\begin{equation}
I_L^0= \frac{Tr(\hat{s_0} \Gamma_1 \hat{s_1} \Gamma_2 \ldots)}
{ \prod_{l=0}^{L-1}s_l^2}
\label{42}
\end{equation}
is the manifestly chiral invariant amplitude corresponding to the
theory without the Wilson term.

The first factor in the eq(\ref{40}) may be expanded in terms of $aq_l$
where $q_l, \quad l=1, \ldots r$ are ``small'' external momenta.
Being integrated over $p, q_{r+1} \ldots q_{L-1}$ the first term in
the eq(\ref{40}) is equal to the manifestly chiral invariant amplitude
(\ref{42}) integrated with the chiral invariant function and
multiplied by a polynomial of external momenta. Such a term is
obviously chiral invariant.

To analyse the next terms we use the relations which follow from the
formal expansion of the eq (\ref{34}) in terms of $n^2$.
\begin{equation}
 \sum_{l=0}^r \frac{B_l}{s_l^2}(1- n^2 \frac{m_l^2}{s_l^2}+ \ldots)=
\label{43}
\end{equation}
$$
=K_0 \prod_{l=0}^r \frac{1}{s_l^2}-n^2(K_1 \prod_{l=0}^r
\frac{1}{s_l^2}+K_0 \prod_{l=0}^r \frac{m_l^2}{s_l^2})
$$
Using these relations we can rewrite the second term in the r.h.s. of
the eq(\ref{40}) in the form
\begin{equation}
 \sum_{k=r+1}^{L-1} \frac{A_ks_k \pi}{m_k^3 \sinh
( \pi s_km_k^{-1})}[K_1 \prod_{l=0}^r \frac{1}{s_l^2}+K_0 \prod_{l=0}^r
\frac{m_l^2}{s_l^4}]
\label{44}
\end{equation}
The first term in brackets is equal to
\begin{equation}
\frac{K_1 \prod_{k=r+1}^{L-1}s_k^2}{K_0}I_L^0
\label{45}
\end{equation}
The factor $K_1K_0^{-1} \prod_ks_k^2$ may be expanded in a power
series in $aq_l, \quad l=1, \ldots r$. Coefficients in this series are
not singular at $p=0$. Therefore being integrated over $p, q_k, \quad
k=r+1, \ldots L-1$ this term is equal to a manifestly chiral invariant
amplitude $I_0$ multiplied by a polynomial in $q_l$. In the second
term in the eq(\ref{44}) we can expand 
$m_l^2s_l^{-4}$ in terms of $aq_l, \quad l=1, \ldots r$. Again it is
easy to see that the coefficients of this expansion are not singular at
$p=0$. Hence being integrated over $p, q_k$ this term produces only
local counterterms.

The next order terms in the eq(\ref{40}) are analyzed in the same way
and lead to the same conclusion. It proves that the derivative over $
\kappa$ of the diagrams under consideration is equal to a chiral
invariant amplitude plus a local gauge invariant polynomial up to
exponentially small corrections. Hence the QCD determinant is chiral
invariant.

These reasonings are easily extended to the diagrams including more
than one fermion loop. A diagram with two fermion loops may be
presented as a product of two diagrams including one fermion loop
integrated with some function depending on the virtual gluon momenta.
For each one loop diagram one has to use the decompositions (\ref{34},
\ref{36}). Multiplying the corresponding expressions and using the
representation
\begin{equation}
\frac{1}{(s_l^2+m_l^2n^2)(s_k^2+m_k^2n^2)}=
\frac{1}{s_k^2m_l^2-m_l^2s_k^2)}( \frac{m_l^2}{s_l^2+m_l^2n^2}-
\frac{m_k^2}{s_k^2+m_k^2n^2})
\label{46}
\end{equation}
one reduces the problem to the one considered above. Diagrams with
arbitrary number of fermion loops are analyzed in the same way.

It completes the proof of the chiral invariance of any gauge invariant
amplitude in our model.

\section{Chiral currents conservation. Anomalies.}

In this section we prove that the currents assosiated with
nonanomalous global chiral symmetries are conserved up to
exponentially small terms.

Let us consider the action (\ref{7}) with the source term for a chiral
current:
\begin {equation}
I_V=I+ \sum_{x, \mu}V_{\mu}^b(x)J_{\mu}^b(x)
\label{47}
\end{equation}
Here $J_{\mu}$ is the axial current which in the continuum case is
equal to
\begin{equation}
J_{\mu}^b(x)= \sum_{n=- \infty, n \neq 0}^{ \infty}\bar{\psi}_n(x) \gamma_5 \gamma_{\mu}T^b \psi_n(x)
\label{48}
\end{equation}
$T^b$ being the generators of the chiral symmetry group. $V_{\mu}$ is
the source term. In the continuum chiral invariant theory this current
is covariantly conserved
\begin{equation}
 \partial_{\mu}J_{\mu}^b+gt^{bcd}J_{\mu}^cV_{\mu}^d=0
\label{49}
\end{equation}
For the lattice action (\ref{47}) including the Wilson term the
corresponding equation is modified. Firstly, the Neuther current has a
contribution from the Wilson term, and secondly, as the Wilson term
breaks the chiral symmetry, the r.h.s. is nonzero. The analogue of the
eq.(\ref{49}) for the action (\ref{47}) may be written as follows
\begin{equation}
 \partial_{\mu}J_{\mu}^b+gt^{bcd}J_{\mu}^cV_{\mu}^d= \sum_{n=- \infty,
\quad n \neq 0}^{\infty}n \sum_{x} \bar{\psi}_n(x) \tilde{W}(x) \psi_n(x)
\label{50}
\end{equation}
where $ \partial$ is a lattice derivative. The r.h.s. of eq.(\ref{50})
is due to the Wilson term and in the quantum theory is responsible for
the chiral anomaly. A particular form of $ \tilde{W}$ is not essential
for our discussion. It is important that the r.h.s. of eq.(\ref{50}) is
bilinear in $ \psi$ and proportional to $n$.

Being integrated with the exponent of the action (\ref{47}) the
eq.(\ref{50}) leads to the chiral Ward identity broken by the Wilson
term
\begin{equation}
 a^{-1} \sin(k_{\mu}a) \Gamma_{ \mu \nu_1 \ldots \nu_n}^{bc_1 \ldots
c_n}(k, q_1 \ldots q_n)+
\label{51}
\end{equation}
$$
+gt^{bcd} \int d^4p \Gamma_{\mu \nu_1 \ldots \nu_n}^{cc_1 \ldots
c_n}(k-p,q_1 \ldots q_n)V_{\mu}^d(p)= \sum_{n=- \infty, \quad n \neq
0
}^{\infty}nW_n
$$
Here $W_n$ stands for the sum of diagrams generated by the r.h.s. of
the eq.(\ref{50}). 

We shall use the differential form of this identity, which is obtained
by differentiating it over $k_{\mu}$ and putting $k=0$:
\begin{equation}
 \Gamma_{\mu \nu_1 \ldots \nu_n}^{bc_1 \ldots c_n}(0,q_1 \ldots q_n)+
\label{52}
\end{equation}
$$
+gt^{bcd} \int d^4p \partial_{\mu} \Gamma_{\mu \nu_1 \ldots \nu_n}^{cc_1
\ldots c_n}(k-p,q_1 \ldots q_n)|_{k=0}V_{\mu}^d(p)= \sum_{n=- \infty, \quad n \neq 0}^{\infty}n
\frac{\partial W_n}{\partial k_{\mu}}|_{k=0}
$$
Here is the important difference between anomalous and nonanomalous
theories. Passing from the eq.(\ref{51}) to the differential identity
(\ref{52}) we assumed that $ \frac{ \partial \Gamma}{\partial
k_{\mu}}$ is not singular at $k=0$. It is not true for anomalous
models: the anomalous triangle diagram is singular at $k=0$. Therefore
the eq.(\ref{52}) is equivalent to the identity (\ref{51}) only if
anomaly is absent.

Now we shall show that the r.h.s. of the eq.(\ref{52}) is
exponentially small. To prove the absence of anomaly it is sufficient
to consider one loop diagrams. The diagrams representing the
r.h.s. of eq.(\ref{52}) are obtained from the QCD amplitudes which we
considered in the previous section by insertions of the additional
vertices $n \frac{ \partial W}{\partial k_{\mu}}|_{k=0}=n \mu$,
corresponding to zero external momentum. In other words they may be
considered as momentum dependent counterterms which shift the Wilson
term $nm \rightarrow n(m+ \mu)$. The sum over $n$ gives for such diagrams the result analogous to the eq.(\ref{24}) 
\begin{equation}
\Pi^L= \int_{- \epsilon a^{-1}}^{ \epsilon a^{-1}}d^4p
[ \frac{A_l(p,q)}{s_l^2}-
\label{53}
\end{equation}
$$
- \sum_{l=0}^{L-1} \frac{ \pi A_l(p,q)}{(m_l+ \mu) \sqrt{s_l^2} \sinh
( \pi \sqrt{s_l^2}(m_l+ \mu)^{-1})}]+c.t.
$$ 
where $c.t.$ stands for local  counterterms. As was
explained above the sum of the first term and local counterterms is
equal to the gauge invariant QCD amplitude without insertions of
$\mu$ and the remaining terms are proportional to $exp \{- \frac{\pi
s_l}{\mu+m_l} \}$. The diagrams with $k$ insertions of $ \mu$ are given by
the $k$-th term of the formal Taylor expansion of this exponent in
terms of $\mu$ near the point $\mu=0$. All these terms are
exponentially small. So we proved that to any nonanomalous global
chiral symmetry corresponds a current which is covariantly conserved
up to exponentially small corrections. At the same time the global
$U(1)$ anomaly is present as it was explained before the differential
Ward identity (\ref{52}) does not hold in this case.

\section{Discussion}

In the previous sections we demonstrated that the lattice QCD
determinant may be presented by a path integral of a local action,
which is free of spectrum degeneracy and reproduces gauge invariant
amplitudes of massless QCD up to exponentially small corrections.
Nonanomalous global chiral symmetries, in particular $SU(2)$ chiral
invariance are also preserved up to exponentially small corrections.
No chirality breaking counterterms are needed and the model provides
automatically $O(a)$ improvement.

Our procedure may be also used in the case when a nonzero bare quark
mass is present. In this case one has to consider the sums of the form
\begin{equation}
 \sum_{n=- \infty, \quad n \neq 0}^{ \infty}
\frac{(-1)^n}{s_l^2+(m_ln+m_0)^2}
\label{54}
\end{equation}
where $m_0$ is the bare quark mass. Representing this sum by the
integral
\begin{equation}
 \int_C \frac{1}{\cos( \pi z)[s_l^2+(m_lz+m_0)^2]}dz
\label{55}
\end{equation}
where the contour $C$ encloses the real axis except for a vicinity of
the point $z=0$, one sees that this integral is equal to
\begin{equation}
\frac{1}{s_l^2+m_0^2}+ O( \exp \{- \pi s_lm_l^{-1} \})
\label{56}
\end{equation}
Of course in this case the chiral current conservation is broken by
the mass term and chirality breaking corrections $ \sim m_0$ are present.

Our effective action includes an infinite series of fields, and for
practical simulations it is important to know how sensitive is the
procedure to cutting this series by some finite number of terms. Fist
of all we note that cutting the series does not spoil the gauge
invariance of the model. So all the arguments using this invariance
are still applicable. Due to the fast convergence of the series we
expect that a small number of terms may provide a good approximation.
This number depends on the value of external momenta $q_l$, which are
relevant for the process under consideration. As we showed above the
integration over momenta bigger than $q_l$ produces local
counterterms. Therefore it is sufficient to study the convergence of
the series for $|p|<|q_l|$. The parameter which determines the value
of the correction term is $ \sqrt{s_l^2}m_l^{-1}=b$. For external
momenta $|qa|<1$ this parameter is $ \geq 2$ (for $\kappa=1$).

The following simple observation may be useful for numerical
simulations. For large values of $b$ the correction term is small,
however to approximate it by a finite series one needs to cut it by
some number $N \gg b$. We recall however that in our scheme the QCD
determinant is presented as a product of the determinants
corresponding to the fields $ \psi_n, \quad - \infty<n< \infty, \quad
n \neq 0$. To improve the convergence for large $b$ one can cut the
series asymmetrically $-N \leq n \leq N+1$. Then for $b \gg N$ and
even $N$ the contribution of the fields with negative $n$ is close to
zero, whereas the contribution of the fields with positive $n$
produces the correct result.

Numerical estimates of the correction term for $qa<1/4, \quad \kappa=1/4$
show that cutting the series by $-2 \leq n \leq 3$ gives the
relative correction $ \sim 0.02$. Of course it is a rather crude and
naive estimate and only real simulations may prove the efficiency of
the method. However it raises a hope that relatively small amount of
fields may be sufficient to have a strong supression of chirality
breaking effects. We recall also that only half of the $ \psi_n$
fields are fermions which is important for simulations with dynamical
quarks. Having in mind that these simulations do not require a fine
tuning of mass parameters and introduction of ``improving'' terms, one
may think that this model provides a competitive method of simulations
in lattice QCD.

 {\bf Acknowledgements.} \\
 This work was done while the author was visiting Center for
Computational Physics of the University of Tsukuba. I am grateful to
Y.Iwasaki, A.Ukawa and all members of CCP for hospitality and fruitful
discussions. I acknowledge stimulating discussions with Y.Kikukawa on
related problems.
This work is supported in part by Russian Basic Research Foundation
under grant   96-01-00551, Presidential grant for the support of
leading scientific schools and INTAS-96-370 $$~$$
 \end{document}